\documentclass[aip,jcp, preprint, groupedaddress]{revtex4-1}%
\usepackage{graphicx}
\usepackage{tabularx}
\usepackage{dcolumn}
\usepackage{longtable}
\usepackage{tensor}
\usepackage{color}
\usepackage{placeins}
\usepackage{bm}
\usepackage{amsmath}
\usepackage{amsfonts}
\usepackage{amssymb}%
\providecommand{\U}[1]{\protect\rule{.1in}{.1in}}
\providecommand{\U}[1]{\protect\rule{.1in}{.1in}}
\providecommand{\U}[1]{\protect\rule{.1in}{.1in}}
\begin{document}
\title{Efficient geometric integrators for nonadiabatic quantum dynamics. II. The
diabatic representation}
\author{Julien Roulet}
\author{Seonghoon Choi}
\author{Ji\v{r}\'{\i} Van\'{\i}\v{c}ek}
\email{jiri.vanicek@epfl.ch}
\affiliation{Laboratory of Theoretical Physical Chemistry, Institut des Sciences et
Ing\'enierie Chimiques, Ecole Polytechnique F\'ed\'erale de Lausanne (EPFL),
CH-1015, Lausanne, Switzerland}
\date{\today}

\begin{abstract}
Exact nonadiabatic quantum evolution preserves many geometric properties of
the molecular Hilbert space. In a companion paper [S. Choi and J.
Van\'{\i}\v{c}ek, 2019], we presented numerical integrators of arbitrary-order
of accuracy that preserve these geometric properties exactly even in the
adiabatic representation, in which the molecular Hamiltonian is not separable
into a kinetic and potential terms. Here, we focus on the separable
Hamiltonian in diabatic representation, where the split-operator algorithm
provides a popular alternative because it is explicit and easy to implement,
while preserving most geometric invariants. Whereas the standard version has
only second-order accuracy, we implemented, in an automated fashion, its
recursive symmetric compositions, using the same schemes as in the companion
paper, and obtained integrators of arbitrary even order that still preserve
the geometric properties exactly. Because the automatically generated
splitting coefficients are redundant, we reduce the computational cost by
pruning these coefficients and lower memory requirements by identifying unique
coefficients. The order of convergence and preservation of geometric
properties are justified analytically and confirmed numerically on a
one-dimensional two-surface model of NaI and a three-dimensional three-surface
model of pyrazine. As for efficiency, we find that to reach a convergence
error of $10^{-10}$, a $600$-fold speedup in the case of NaI and a $900$-fold
speedup in the case of pyrazine are obtained with the higher-order
compositions instead of the second-order split-operator algorithm. The
pyrazine results suggest that the efficiency gain survives in higher dimensions.

\end{abstract}
\maketitle

\graphicspath{{./figures/}
{C:/Users/Jiri/Dropbox/Papers/Chemistry_papers/2019/Integrators_diabatic/Integrators_diabatic_v24/figures/}
{"d:/Group Vanicek/Desktop/Integrators_diabatic_v18/figures/"}}

\section{\label{sec:intro}Introduction}

The celebrated Born--Oppenheimer
approximation\cite{Born_Oppenheimer:1927,book_Heller:2018} assumes the
separability of the nuclear and electronic motions in a molecule, and provides
an appealing picture of independent electronic potential energy surfaces.
However, many important processes in nature\cite{Mokhtari_Zewail:1990} can
only be described by considering nonadiabatic couplings between these
Born-Oppenheimer
surfaces.\cite{book_Baer:2006,book_Nakamura:2012,book_Takatsuka:2015,Bircher_Rothlisberger:2017}
To investigate such processes, one can abandon the Born-Oppenheimer
representation and treat electrons and nuclei
explicitly,\cite{Shin_Metiu:1995,Albert_Engel:2016,Matyus:2019} use an exact
factorization\cite{Abedi_Gross:2010,Cederbaum:2008} of the molecular
wavefunction, or, determine which Born-Oppenheimer states are coupled
strongly\cite{Zimmermann_Vanicek:2010,Zimmermann_Vanicek:2012} and then solve
the time-dependent Schr\"{o}dinger equation with a nonadiabatically coupled
molecular Hamiltonian; below, we will only consider the third and most common strategy.

In a companion paper\cite{Choi_Vanicek:2019} (which will be referred to as
Paper I), we surveyed several algorithms for the nonadiabatic quantum
dynamics, applicable to higher dimensions, including Gaussian basis
methods,\cite{Ben-Nun_Martinez:2000,
Curchod_Martinez:2018,Shalashilin_Child:2001a,Makhov_Shalashilin:2017,
Worth_Burghardt:2004,Richings_Lasorne:2015} variations of the
multiconfigurational time-dependent Hartree (MCTDH)
method,\cite{Meyer_Cederbaum:1990,Worth_Burghardt:2008,Wang_Thoss:2003} and
sparse-grid methods.\cite{Avila_Carrington:2017,book_Lubich:2008} There are
situations, however, in which the wavepacket spreads over large parts of the
available Hilbert space, and then time-independent basis sets or full-grid
methods can become more efficient.

As for the molecular Hamiltonian used in nonadiabatic simulations, the
\emph{ab initio} electronic structure methods typically yield the
\emph{adiabatic} potential energy surfaces, which are nonadiabatically coupled
via momentum couplings. However, in the regions of conical
intersections,\cite{Worth_Cederbaum:2004,Domcke_Yarkony:2012} the
Born-Oppenheimer surfaces become degenerate, and the nonadiabatic couplings
diverge. To avoid associated problems, it is convenient to use the diabatic
representation, in which the divergent momentum couplings are replaced with
well-behaved coordinate couplings. Although exact diabatization is only
possible in systems with two electronic states and one nuclear degree of
freedom,\cite{Mead_truhlar:1982} there exist more general, 
approximate diabatization
procedures,\cite{Granucci_Toniolo:2001,Nakamura_Truhlar:2002,Baer:2002}
starting with the vibronic coupling Hamiltonian
model.\cite{Koppel_Cederbaum:1984} Another benefit of the diabatic
representation is that it separates the Hamiltonian into a sum of kinetic
energy, depending only on nuclear momenta, and potential energy, depending
only on nuclear coordinates, which makes it possible to propagate the
molecular wavefunction with the split-operator
algorithm.\cite{Feit_Steiger:1982, book_Lubich:2008, book_Tannor:2007} The
split-operator algorithm is explicit, easy to implement, and, in addition, it
is an example of a geometric
integrator\cite{book_Hairer_Wanner:2006,book_Leimkuhler_Reich:2004} because,
similarly to the integrators discussed in Paper I,\cite{Choi_Vanicek:2019} it
conserves \emph{exactly }many invariants of the exact solution, regardless of
the convergence error of the wavefunction itself. Geometric integrators in
general acknowledge special properties of the Schr\"{o}dinger equation which
differentiate it from other differential equations. Using these integrators
can be likened to using a well-fitting screw-driver instead of a hammer to
attach a screw. Note that the integrators for nonseparable Hamiltonians,
presented in Paper I, are also geometric and, clearly, still applicable to the
separable Hamiltonian in the diabatic representation, but the split-operator
algorithm is expected to be more efficient because it is explicit.

The standard, second-order split-operator algorithm\cite{Feit_Steiger:1982} is
unitary, symplectic, stable, symmetric, and time-reversible, regardless of the
size of the time step. However, to obtain highly accurate results, the
standard algorithm requires using a small time step, because it has only
second-order accuracy. There exist much more efficient algorithms, such as the
short-iterative Lanczos algorithm,\cite{Lanczos:1950, Park_Light:1986,
Kuleff_Cederbaum:2005} which has an exponential convergence with respect to
the time step, and also conserves the norm and energy, but not the inner
product (because it is nonlinear) and other geometric properties.

To address the low accuracy of the second-order split-operator algorithm and
the nonconservation of geometric properties by other more accurate methods,
various higher-order split-operator integrators have been
introduced,\cite{Forest_Ruth:1990,Suzuki:1990,Yoshida:1990,Bandrauk_Shen:1991}
some of which allow complex time
steps\cite{Bandrauk_Shen:1991,Bandrauk_Dehghanian:2006,Prosen_Pizorn:2006} or
commutators of the kinetic and potential energies in the
exponent,\cite{Raedt:1987,Suzuki:1995,Chin_Chen:2002} thus reducing the number
of splitting steps. Here we explore one type of higher-order integrators,
designed for nonadiabatic dynamics in the diabatic basis, which we have
implemented using the recursive triple-jump\cite{Suzuki:1990,Yoshida:1990} and
Suzuki-fractal,\cite{Suzuki:1990} as well as several non-recursive,
\textquotedblleft optimal\textquotedblright\ compositions of the second-order
split-operator algorithm. While the recursive compositions permit an automated
generation of integrators of arbitrary even order in the time
step,\cite{Yoshida:1990, McLachlan:1995, Suzuki:1990,
book_Leimkuhler_Reich:2004,book_Hairer_Wanner:2006,Wehrle_Vanicek:2011} the
efficiency of higher-order algorithms is sometimes questioned because the
number of splitting steps grows exponentially with the order of accuracy, and,
consequently, so does the computational cost of a single time step. Motivated
by this dilemma, we have explored the convergence and efficiency of the
higher-order compositions using a one- and three-dimensional systems,
concluding that, despite the increasing number of splittings, the higher-order
methods become the most efficient if higher accuracy of the solution is
required, and that this gain in efficiency survives in higher dimensions. We
have also confirmed that all composed methods are unitary, symplectic, stable,
symmetric, and time-reversible. A final benefit of the higher-order methods is
the simple, abstract, and general implementation of the compositions of the
second-order split-operator algorithm; indeed, even this \textquotedblleft
elementary\textquotedblright\ method is a composition of simpler, first-order
algorithms.\cite{book_Lubich:2008, book_Tannor:2007}

One of the only challenges of implementing the split-operator algorithm for
nonadiabatic dynamics in the diabatic representation is the exponentiation of
the potential energy operator, which is nondiagonal in the electronic degrees
of freedom (in contrast to the diagonal kinetic energy operator). We,
therefore, explored several methods for the exponentiation of nondiagonal matrices.

The main disadvantage of the split-operator algorithm and its compositions is
that their use is restricted to separable Hamiltonians. To compare them with
the integrators from Paper I, we cannot use the adiabatic representation, but
instead must perform the comparison in the diabatic representation, where the
compositions of the explicit split-operator algorithm are, as expected, much
more efficient than the more generally applicable
compositions\cite{Choi_Vanicek:2019} of the implicit trapezoidal rule (the
Crank-Nicolson method\cite{Crank_Nicolson:1947,McCullough_Wyatt:1971}) from
Paper I. Nevertheless, the comparison serves as a higher-dimensional test of
integrators from Paper I and confirms that, in contrast to the split-operator
compositions, the integrators from Paper I conserve also the energy exactly.

The remainder of this paper is organized as follows: In Sec.~\ref{sec:theory},
after reviewing the geometric properties of the exact evolution operator, we
discuss the lack of symmetry and time-reversibility in the first-order
split-operator algorithms and the recovery of these properties in the
symmetric compositions. Next, we describe several strategies for reducing the
computational cost and memory requirements by pruning redundant splitting
coefficients generated automatically by the symmetric compositions. After
presenting the dynamic Fourier method for its ease of implementation and the
exponential convergence with the grid density, we briefly discuss the
molecular Hamiltonian in diabatic representation. In Section~\ref{sec:results}%
, the convergence properties and conservation of geometric invariants by
various methods are analyzed numerically on a one-dimensional two-surface
model\cite{Engel_Metiu:1989} of NaI and a three-dimensional three-surface
model of pyrazine,\cite{Stock_Woywod:1995} both in the diabatic
representation. Section~\ref{sec:conclusion} concludes the paper.

\section{\label{sec:theory}Theory}

\subsection{\label{subsec:exactprop}Geometric properties of the exact
evolution operator}

The time-dependent Schr\"{o}dinger equation
\begin{equation}
i\hbar\frac{d\psi(t)}{dt}=\hat{H}\psi(t) \label{eq:tdse}%
\end{equation}
with a time-independent Hamiltonian $\hat{H}$ and initial condition $\psi(0)$
has the formal solution $\psi(t)=\hat{U}(t)\psi(0)$, where $\hat{U}(t)$ is the
evolution operator. While in Paper I, we considered general Hamiltonian
operators $\hat{H}\equiv H(\hat{q},\hat{p})$, here we require that the
Hamiltonian be separable as%
\begin{equation}
\hat{H}\equiv\hat{T}+\hat{V}\equiv T(\hat{p})+V(\hat{q})
\label{eq:H_separable}%
\end{equation}
into a sum of kinetic and potential energies, which depend, respectively, only
on the momentum $\hat{p}$ and position $\hat{q}$ operators.

The exact evolution operator
\begin{equation}
\hat{U}(t)=e^{-i\hat{H}t/\hbar}=e^{-i[T(\hat{p})+V(\hat{q})]t/\hbar}
\label{eq:exact_evol_op}%
\end{equation}
is linear, unitary, symplectic, symmetric, time-reversible, stable, and
conserves the norm, inner product, and energy. Because these properties are
desirable also in approximate numerical evolution operator $\hat
{U}_{\text{appr}}(t)$, let us define them briefly.

An operator $\hat{U}$ is said to \emph{preserve the norm} if $\Vert\hat{U}%
\psi\Vert=\Vert\psi\Vert$ for all $\psi$, and to \emph{preserve the inner
product} if $\langle\hat{U}\psi|\hat{U}\phi\rangle=\langle\psi|\phi\rangle$
for all $\psi$ and $\phi$. For linear operators $\hat{U}$, these two
properties are equivalent, whereas for general, possibly nonlinear operators,
conservation of the inner product implies linearity\cite{book_Halmos:1942} and
hence the conservation of norm, but norm conservation implies neither
linearity nor conservation of the inner product. An operator $\hat{U}$ is said
to be \emph{unitary} if $\hat{U}^{\dagger}=\hat{U}^{-1}$, where $\hat{U}%
^{\dag}$ is the Hermitian adjoint. An operator $\hat{U}$ is called
\emph{symplectic} if $\omega(\hat{U}\psi,\hat{U}\phi)=\omega(\psi,\phi)$,
where $\omega(\psi,\phi)$ is a \emph{symplectic two-form}, i.e., a
nondegenerate skew-symmetric bilinear form. We will only consider the
symplectic two-form defined as\cite{book_Lubich:2008} $\omega(\psi
,\phi):=-2\hbar\mathrm{Im}\langle\psi|\phi\rangle$, which is, obviously,
conserved, if the inner product is. $\hat{U}$ is said to \emph{conserve
energy} if $\langle\hat{H}\rangle_{\hat{U}\psi}=\langle\hat{H}\rangle_{\psi}$,
where $\langle\hat{A}\rangle_{\psi}:=$ $\langle\psi|\hat{A}|\psi\rangle$
denotes the expectation value of operator $\hat{A}$ in the state $\psi$.
Finally, an \emph{adjoint} $\hat{U}(t)^{\ast}$ of an evolution operator
$\hat{U}(t)$ is defined as $\hat{U}(t)^{\ast}:=\hat{U}(-t)^{-1}$. An evolution
operator is said to be \emph{symmetric}
if\cite{book_Leimkuhler_Reich:2004,book_Hairer_Wanner:2006} $\hat{U}(t)^{\ast
}=\hat{U}(t)$ and \emph{time-reversible}
if\cite{book_Leimkuhler_Reich:2004,book_Hairer_Wanner:2006} $\hat{U}%
(-t)\hat{U}(t)\psi=\psi$. For the definition of stability and a more detailed
presentation and discussion of other properties, see Sec.~II~A of Paper I.

\subsection{\label{subsec:lossprop}First-order split-operator methods}

In approximate propagation methods, the state at time $t+\Delta t$ is obtained
from the state at time $t$ using the relation%
\[
\psi(t+\Delta t)=\hat{U}_{\mathrm{appr}}(\Delta t)\psi(t)
\]
where $\hat{U}_{\mathrm{appr}}(\Delta t)$ is an approximate time evolution
operator and $\Delta t$ the numerical time step. Depending on the order of
kinetic and potential propagations, the approximate evolution operator is
\begin{equation}
\hat{U}_{\text{VT}}(\Delta t):=e^{-\frac{i}{\hbar}\Delta t{\hat{V}}}%
e^{-\frac{i}{\hbar}\Delta t{\hat{T}}} \label{eq:so_VT}%
\end{equation}
in the \emph{VT split-operator algorithm} and
\begin{equation}
\hat{U}_{\text{TV}}(\Delta t):=e^{-\frac{i}{\hbar}\Delta t\hat{T}}e^{-\frac
{i}{\hbar}\Delta t\hat{V}} \label{eq:so_TV}%
\end{equation}
in the \emph{TV split-operator algorithm}. Both $\hat{U}_{\text{VT}}$ and
$\hat{U}_{\text{TV}}$ are unitary, symplectic, stable, but only first-order in
the time step $\Delta t$. Neither method conserves energy because neither
evolution operator commutes with the Hamiltonian. Neither method is symmetric;
in fact, they are adjoints of each other. Hence, neither method is
time-reversible. These properties are justified in Appendix~\ref{appendixa}
and summarized in Table~\ref{tab:properties}.

Although the first-order split-operator algorithms are not time-reversible,
composing them in a specific way leads to time-reversible integrators of
arbitrary order of accuracy in the time step.

\begin{table}
[pbh]%
\caption{Geometric properties and computational cost of the first-order and
recursively composed second-order split-operator (SO) algorithms. Cost (here
before speedup by pruning splitting coefficients) is measured by the number of
fast Fourier transforms required per time step (see Sec.~\ref{sec:dyn_fourier}). $n$ is the number of recursive compositions and $C$ the total number of
composition steps per time step ($C = 3^{n}$ for the triple
jump\cite{Yoshida:1990, Suzuki:1990}, $C = 5^{n}$ for Suzuki's
fractal\cite{Suzuki:1990}). $+$ or $-$ denotes that the geometric property of
the exact evolution operator is or is not preserved.}\label{tab:properties}%
\begin{ruledtabular}
\begin{tabular}{lccccccccc}
\multicolumn{1}{c}{}     Method         & \multicolumn{1}{c}{Order} &
\multicolumn{1}{c}{Unitary} & \multicolumn{1}{c}{Symplectic} &
\multicolumn{1}{c}{Commutes} & \multicolumn{1}{c}{Energy} &
\multicolumn{1}{c}{Symm-} &
\multicolumn{1}{c}{Time-} & \multicolumn{1}{c}{Stable} &
\multicolumn{1}{c}{Cost}\\
\multicolumn{1}{c}{}              & \multicolumn{1}{c}{} & \multicolumn{1}{c}{}
& \multicolumn{1}{c}{} &
\multicolumn{1}{c}{with $\hat{H}$} & \multicolumn{1}{c}{cons.} &
\multicolumn{1}{c}{etric} &
\multicolumn{1}{c}{reversible} & \multicolumn{1}{c}{} & \\ \hline
$1^{\mathrm{st}}$ order SO  & 1                         & $+$
& $+$                              & $-$
& $-$                                  & $-$                            & $-$
& $+$ &
$2$
\\
$2^{\mathrm{nd}}$ order SO & 2$(n+1)$                      & $+$
& $+$                              & $-$
& $-$                                  & $+$                            & $+$
& $+$  &
$2 C$
\\
\end{tabular}
\end{ruledtabular}

\end{table}

\subsection{\label{subsec:compprop}Recovery of geometric properties by
composed methods}

Composing the two first-order split-operator algorithms, each for a time step
$\Delta t/2,$ yields a symmetric second-order method.\cite{Feit_Steiger:1982}
Depending on the order of composition, one obtains either the \emph{VTV
algorithm}%
\begin{equation}
\hat{U}_{\text{VTV}}(\Delta t):=\hat{U}_{\text{VT}}(\Delta t/2)\hat
{U}_{\text{TV}}(\Delta t/2), \label{eq:so_VTV}%
\end{equation}
or \emph{TVT algorithm}%
\begin{equation}
\hat{U}_{\text{TVT}}(\Delta t):=\hat{U}_{\text{TV}}(\Delta t/2)\hat
{U}_{\text{VT}}(\Delta t/2). \label{eq:so_TVT}%
\end{equation}
Both are explicit, unitary, symplectic, stable, symmetric, and
time-reversible, regardless of the size of the time step. Neither evolution
operator commutes with the Hamiltonian and, therefore, neither method
conserves energy exactly. These properties are again justified in
Appendix~\ref{appendixa} and summarized in Table~\ref{tab:properties}.

\subsection{\label{subsec:composition}Symmetric composition schemes for
symmetric methods}

As discussed in Paper I, composing any symmetric second-order method (such as
one of those of Sec.~\ref{subsec:compprop}) with appropriately chosen time
steps leads to symmetric integrators of arbitrary order of
accuracy.\cite{book_Leimkuhler_Reich:2004,book_Hairer_Wanner:2006,Suzuki:1990,Yoshida:1990}
More precisely, there are a natural number $M$ and real numbers $\gamma_{n}$,
$n=1,\ldots,M$, called \emph{composition coefficients}, such that $\gamma
_{1}+\cdots+\gamma_{M}=1$ and such that for any symmetric evolution operator
$\hat{U}_{p}(\Delta t)$ of an even order $p$, composing this symmetric
evolution operator with coefficients $\gamma_{n}$ yields a symmetric
integrator of order $p+2$:%
\[
\hat{U}_{p+2}(\Delta t):=\hat{U}_{p}(\gamma_{M}\Delta t)\cdots\hat{U}%
_{p}(\gamma_{1}\Delta t).
\]

The simplest composition schemes (see Fig.~2 of
Ref.~\onlinecite{Choi_Vanicek:2019}) are the triple
jump\cite{Cruetz_Gocksch:1989,Forest_Ruth:1990,Suzuki:1990,Yoshida:1990} with
$M=3$, and Suzuki's fractal\cite{Suzuki:1990} with $M=5$. Both are
\emph{symmetric compositions}, meaning that $\gamma_{M+1-n}=\gamma_{n}$.
Because larger time steps can be used for calculations using Suzuki's fractal,
this composition is sometimes more efficient than the triple-jump composition,
despite requiring more composition steps (see
Ref.~\onlinecite{Choi_Vanicek:2019} for a numerical example). For specific
orders of convergence, more efficient non-recursive composition schemes exist
and will be referred to as \textquotedblleft optimal.\textquotedblright\ These
were implemented according to Kahan and Li\cite{Kahan_Li:1997} for the
$6^{\mathrm{th}}$ and $8^{\mathrm{th}}$ orders, and according to Sofroniou and
Spaletta\cite{Sofroniou_Spaletta:2005} for the $10^{\mathrm{th}}$ order (see
Sec.~II~D of Paper I for more details about composition methods).

\subsection{\label{subsec:composition_of_so_algorithms}Compositions of
split-operator algorithms}

The split-operator algorithm is applicable if the Hamiltonian $\hat{H}$ can be
written as a sum
\begin{equation}
\hat{H}=\hat{A}+\hat{B} \label{eq:separable_hamiltonian}%
\end{equation}
of operators $\hat{A}$ and $\hat{B}$ with evolution operators, $\hat{U}%
_{\hat{A}}(t)=\exp(-it\hat{A}/\hbar)$ and $\hat{U}_{\hat{B}}(t)=\exp
(-it\hat{B}/\hbar)$, whose actions on $\psi$ can be evaluated exactly. A
general split-operator evolution operator can be expressed as
\[
\hat{U}_{\hat{A}+\hat{B}}^{\mathrm{SO}}(\Delta t)=\hat{U}_{\hat{B}}%
(b_{N}\Delta t)\hat{U}_{\hat{A}}(a_{N}\Delta t)\cdots\hat{U}_{\hat{B}}%
(b_{1}\Delta t)\hat{U}_{\hat{A}}(a_{1}\Delta t),
\]
where $N$ is the number of splitting steps, and $a_{j}$ and $b_{j}$ are the
splitting coefficients associated with the operators $\hat{A}$ and $\hat{B}$.
These coefficients in general satisfy the identity $\sum_{j=1}^{N}a_{j}%
=\sum_{j=1}^{N}b_{j}=1$, and are $a_{1}=b_{1}=1$ for the first-order VT and
TV\ algorithms\cite{Trotter:1959} and
\begin{equation}
a_{1}=a_{2}=\frac{1}{2},\qquad b_{1}=1,\qquad b_{2}=0
\label{eq:second_order_so}%
\end{equation}
for the second-order VTV or TVT algorithms.\cite{Strang:1968}

\begin{figure}
[pth]%
\includegraphics[scale=1.0]{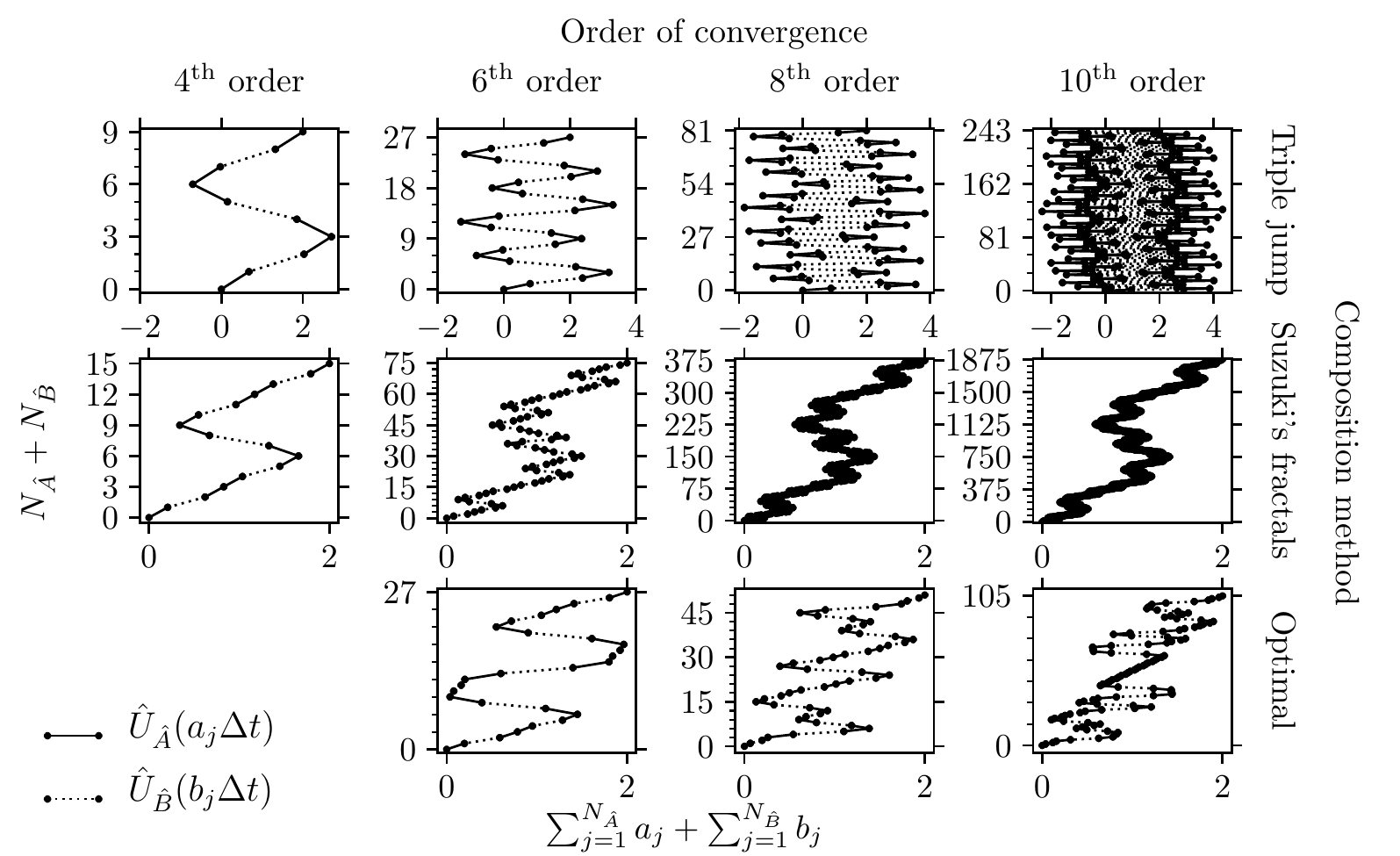}\caption{Split-operator
algorithms composed by the recursive (triple jump and Suzuki's fractal) and
nonrecursive \textquotedblleft optimal\textquotedblright\ composition schemes
shown in Fig.~$2$ of Paper~I.\cite{Choi_Vanicek:2019} In other 
words, each elementary method
$\hat{U}(\gamma_{n}\Delta t)$
(solid line segment in Fig.~$2$ of Paper I) is replaced by a second-order split-operator algorithm
$\hat{U}_{\hat{A}}(\gamma_{n}\Delta t/2)\hat{U}_{\hat{B}}(\gamma_{n}\Delta t)
\hat{U}_{\hat{A}}(\gamma_{n}\Delta t/2)$,
represented here by a triple of consecutive solid, dotted, and solid line segments.
Solid line segments
represent $\hat{U}_{\hat{A}}(\gamma_{n} \Delta t/2)$, whereas the dotted line segments represent
$\hat{U}_{ \hat{B}}(\gamma_{n} \Delta t)$. $N_{\hat{O}}$ is the number of 
actions of
$\hat{U}_{\hat{O}}$ on $\psi$. }\label{fig:triple_suzuki_optimal_so}
\end{figure}

Because the second-order split-operator algorithm\cite{Strang:1968} is
symmetric, it can be composed by any of the composition schemes discussed in
Sec.~\ref{subsec:composition}. For example, the splitting coefficients of a
fourth-order method are
\begin{align}
a_{1}  &  =a_{2}=\frac{1}{2(2-2^{1/3})},\qquad a_{3}=-\frac{2^{1/3}%
}{2(2-2^{1/3})},\nonumber\\
b_{1}  &  =\frac{1}{2-2^{1/3}},\qquad b_{3}=-\frac{2^{1/3}}{2-2^{1/3}},\qquad
b_{2}=b_{6}=0 \label{eq:triple_jump_fourth_order_so}%
\end{align}
with $N=6$ if the triple-jump composition scheme is used, and
\begin{align}
a_{1}  &  =a_{2}=a_{3}=a_{4}=\frac{1}{2(4-4^{1/3})},\quad a_{5}=-\frac
{4^{1/3}}{2(4-4^{1/3})},\nonumber\\
b_{1}  &  =b_{3}=\frac{1}{4-4^{1/3}},\quad b_{2}=b_{4}=b_{10}=0,\quad
b_{5}=-\frac{4^{1/3}}{4-4^{1/3}} \label{eq:suzukis_fractals_fourth_order_so}%
\end{align}
with $N=10$ if Suzuki's fractal is used instead. The remaining coefficients
are obtained from symmetry as
\begin{equation}
a_{N-j+1}=a_{j},\quad b_{N-j}=b_{j}.
\label{eq:symmetric_splitting_coefficients}%
\end{equation}
Both composition procedures can be applied recursively to obtain higher-order
split-operator algorithms.
These as well as the optimally composed algorithms of up to the tenth order are 
represented
pictorially in Fig.~\ref{fig:triple_suzuki_optimal_so}. 

All compositions of the second-order VTV or TVT split-operator algorithms are
unitary, symplectic, and stable; all symmetric compositions are symmetric and,
therefore, time-reversible. The proof of this statement is a special case of
the general proof of a corresponding theorem for the composition of geometric
integrators in Paper I.

\subsection{\label{subsec:pruning_coefficients}Pruning splitting coefficients}

\begin{figure}
[pth]%
\includegraphics[scale=1.0]{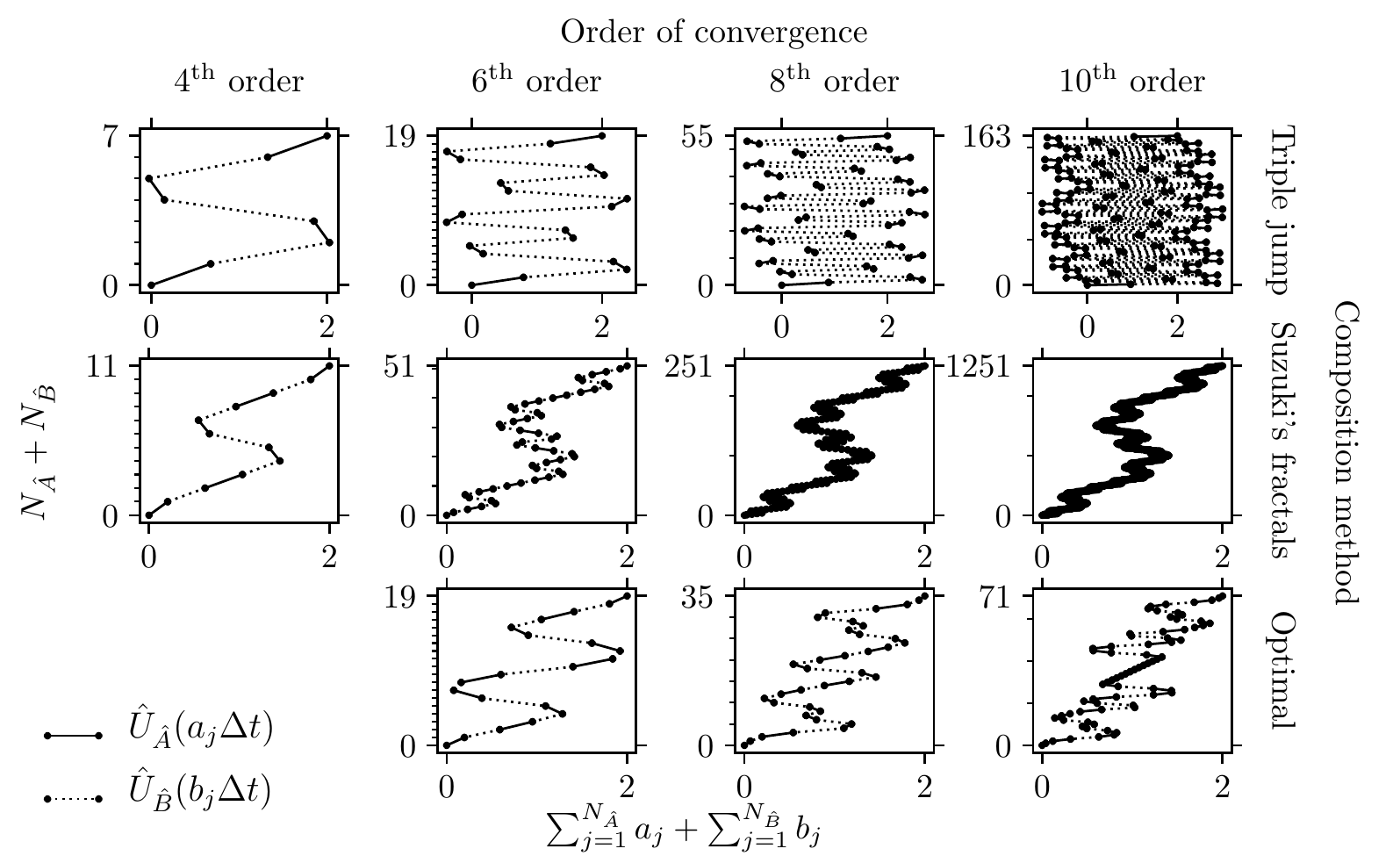}\caption{Composed
split-operator algorithms from Fig.~\ref{fig:triple_suzuki_optimal_so} after
removing zero splitting coefficients and merging adjacent
coefficients, i.e., after each two adjacent solid line segments 
representing
$\hat{U}_{\hat{A}}(\gamma_{n+1} \Delta t/2) \hat{U}_{\hat{A}}(\gamma_{n} \Delta t/2)$
in Fig.~\ref{fig:triple_suzuki_optimal_so} are merged into a single solid line segment representing
$\hat{U}_{\hat{A}}((\gamma_{n} +\gamma_{n+1})\Delta 
t/2)$.}\label{fig:triple_suzuki_optimal_so_pruned}%

\end{figure}

Many $b_{j}$ coefficients of the higher-order integrators obtained by
recursive composition of the second-order split-operator algorithm are zero
[for an example, see Eqs.~(\ref{eq:triple_jump_fourth_order_so}%
)~and~(\ref{eq:suzukis_fractals_fourth_order_so})]. The computational time can
be reduced by \textquotedblleft pruning,\textquotedblright\ i.e., removing the
splitting steps corresponding to $b_{j}=0$ and merging the consecutive actions
of $\hat{U}_{\hat{A}}(a_{j}\Delta t)$ and $\hat{U}_{\hat{A}}(a_{j+1}\Delta
t)$. If $b_{j}=0$ and $j\neq N$, the splitting coefficients are modified as
\begin{align}
\tilde{b}_{k}  &  =b_{k+1},\quad\text{for}\quad j\leq k\leq N-1,\nonumber\\
\tilde{a}_{j}  &  =a_{j}+a_{j+1},\nonumber\\
\tilde{a}_{k}  &  =a_{k+1},\quad\text{for}\quad j+1\leq k\leq
N-1,\nonumber\label{eq:merging_splitting_coeff_b_zero}\\
\tilde{N}  &  =N-1,
\end{align}
in order to merge the $j^{\mathrm{th}}$ and $(j+1)^{\mathrm{th}}$ steps. The
composed methods after the merge are exhibited in
Fig.~\ref{fig:triple_suzuki_optimal_so_pruned} and the reduction in the number
$N$ of splitting steps, which measures the computational cost, is summarized
in Table~\ref{tab:N_SO_pruned_unq}.

For a time-independent separable Hamiltonian, one can either precompute and
store the evolution operators, $\hat{U}_{\hat{A}}(a_{j}\Delta t)$ and $\hat
{U}_{\hat{B}}(b_{j}\Delta t)$, or compute them on the fly. The former approach
is more memory intensive than the latter, which does not store any evolution
operators, but the computational time is reduced since the evolution operators
are only computed once at initialization. To alleviate the memory requirement
of the former approach, one can exploit the repetition of certain splitting
coefficients, which is obvious from Eqs.~(\ref{eq:triple_jump_fourth_order_so}%
)~and~(\ref{eq:suzukis_fractals_fourth_order_so}) and
Fig.~\ref{fig:triple_suzuki_optimal_so_pruned}. If either $\hat{A}$ or
$\hat{B}$ is time-dependent, it is always beneficial to compute the
corresponding evolution operator pertaining to the time-dependent operator on
the fly because no reduction in computational time is possible by precomputing
the evolution operators.

The effort spent in searching for repeated coefficients is reduced if the
symmetries of the composition scheme and of the elementary method are
exploited [see Eq.~(\ref{eq:symmetric_splitting_coefficients})]. The repeated
coefficients are then identified from only half of the original coefficients
$a_{j}$ and $b_{j}$.

Once identified, only the unique evolution operators $\hat{U}_{\hat{A}}%
(a_{j}^{\mathrm{unq}}\Delta t)$ and $\hat{U}_{\hat{B}}(b_{j}^{\mathrm{unq}%
}\Delta t)$ are stored in arrays of lengths $N_{a}^{\mathrm{unq}}$ and
$N_{b}^{\mathrm{unq}}$, together with the information when to apply them,
stored in integer arrays $I^{a}$ and $I^{b}$ of length $N$, containing the
indices in unique coefficient arrays, i.e.,
\begin{equation}
1\leq I_{j}^{a}\leq N_{a}^{\mathrm{unq}},\qquad1\leq I_{j}^{b}\leq
N_{b}^{\mathrm{unq}}. \label{eq:range_index_unq}%
\end{equation}
Exploiting the repeated coefficients, the number of stored evolution operators
reduces from $2N$ to $N_{a}^{\mathrm{unq}}+N_{b}^{\mathrm{unq}}$ (see
Table~\ref{tab:N_SO_pruned_unq}).

\begin{table}
[pbh]\caption{Computational cost and memory requirement of the composed
split-operator algorithms before and after
pruning (i.e., removing zero coefficients and merging adjacent coefficients) and
identifying repeated coefficients.
The computational cost is measured by
$N_{\hat{A}} + N_{\hat{B}}$, where $N_{\hat{O}}$ is the number of actions of
$\hat{U}_{\hat{O}}$ on
the wavepacket.
The memory requirement before and after pruning is 
$N_{\hat{A}}+N_{\hat{B}}$,
and after
identifying repeated coefficients decreases to $N^{\text{unq}}_{a} +
N^{\text{unq}}_{b}$.} \label{tab:N_SO_pruned_unq} \begin{ruledtabular}
\begin{tabular}{cccccc}
Composition &Order&$N_{\hat{A}} + N_{\hat{B}}$ & $N_{\hat{A}} + N_{\hat{B}}$ &
$N_{a}^{\mathrm{unq}}$  &$N_{b}^{\mathrm{unq}}$  \\
method &  &
before merge \footnote{$N_{\hat{A}} = 2 N_{\hat{B}}$  for order 
$\ge 2$.} &
after merge \footnote{$N_{\hat{A}} = N_{\hat{B}} + 1$ for order 
$\ge 2$.} &   &  \\ \hline
Elementary & 1 & 2  & 2  & 1  & 1 \\
methods & 2 & 3  & 3  & 1 & 1 \\ \hline
& 4 & 9  & 7 & 2 &2  \\
Triple &6  & 27  & 19 & 4 &4  \\
jump &8  & 81 & 55 &8  &8  \\
&10  & 243 & 163 &16  &16  \\ \hline
& 4  & 15 & 11 & 3 & 2 \\
Suzuki's &6  & 75 &51  & 6 & 4 \\
fractal &8  & 375 &251  &12  & 8 \\
&10  & 1875 & 1251 & 24 & 16 \\ \hline
&6  & 27 & 19  &5  & 5 \\
Optimal &8  & 51 & 35 & 9 & 9 \\
&10  & 105  & 71 & 18 & 18
\end{tabular}
\end{ruledtabular}

\end{table}

\subsection{\label{sec:dyn_fourier}Dynamic Fourier method}

To propagate a wavepacket $\psi(t)$ with any split-operator algorithm (see
Secs.~\ref{subsec:lossprop}--\ref{subsec:composition}), only the actions of
the kinetic ($\hat{U}_{\hat{T}})$ and potential ($\hat{U}_{\hat{V}}$)
evolution operators on $\psi(t)$ are required, where
\[
\hat{U}_{\hat{T}}(\Delta t):=e^{-i\Delta tT(\hat{p})/\hbar}\text{ and }\hat
{U}_{\hat{V}}(\Delta t):=e^{-i\Delta tV(\hat{q})/\hbar}.
\]
Since $\hat{U}_{\hat{T}}$ and $\hat{U}_{\hat{V}}$ are diagonal in the momentum
and position representations, respectively, their action on $\psi(t)$ is easy
to evaluate in the appropriate representation. This is the main idea of the
dynamic Fourier method,\cite{Feit_Steiger:1982, Kosloff_Kosloff:1983a,
Kosloff_Kosloff:1983, book_Tannor:2007} in which the representation of
$\psi(t)$ is repeatedly changed, as needed, via the fast Fourier transform
(for more details, see Sec.~II~E of Paper I).

In the numerical examples below, the Fourier transform was performed using the
Fastest Fourier Transform in the West 3 (FFTW3)
library.\cite{Frigo_Johnson:2005} Although its accuracy is sufficient for most
applications, small deviations from unitarity, which were due to the high
number of repeated application of the forward and backward Fourier transforms,
affected the most converged calculations. To reduce the nonunitarity, we used
the long-double instead of the default double precision version of FFTW3.

\subsection{Molecular Hamiltonian in the diabatic basis\label{sec:mol_ham}}

The molecular Hamiltonian in the diabatic basis can be expressed as
\begin{equation}
\hat{\mathbf{H}}=\frac{1}{2}\hat{p}^{T}\cdot m^{-1}\cdot\hat{p}\,\mathbf{1}%
+\mathbf{V}(\hat{q}),\label{eq:diab_mol_ham}%
\end{equation}
where $m$ is the diagonal $D\times D$ nuclear mass matrix, $D$ the number of
nuclear degrees of freedom, and $\mathbf{V}$ the potential energy. In
Eq.~(\ref{eq:diab_mol_ham}), the dot $\cdot$ denotes the matrix product in
nuclear $D$-dimensional vector space, the hat $\hat{}$ represents a nuclear
operator, and the \textbf{bold} font indicates an electronic operator, i.e.,
an $S\times S$ matrix, where $S$ is the number of included electronic states.
Using the dynamic Fourier method, each evaluation of the action of the pair
$\mathbf{\hat{U}}_{\mathbf{\hat{V}}}(t_{V})$ and $\mathbf{\hat{U}%
}_{\mathbf{\hat{T}}}(t_{T})$ on a molecular wavepacket $\bm{\psi}(t)$, which
now becomes an $S$-component vector of nuclear wavepackets (one on each
surface), involves two changes of the wavepacket's representation. The
above-mentioned nonunitarity of the solution, partially due to the numerical
implementation of the FFT algorithm, was made worse by the matrix exponential
required for evaluating the potential evolution operator $\mathbf{\hat{U}%
}_{\mathbf{\hat{V}}}(t_{V})$, which contains offdiagonal couplings between the
electronic states. Although we tried different approaches for matrix
exponentiation, including Pad\'{e}
approximants\cite{book_Golub_Van_Loan:1996,Sidje:1998} and exponentiating a
diagonal matrix obtained with the QR
decomposition\cite{book_Golub_Van_Loan:1996,book_Anderson_Sorensen:1999} or
with the Jacobi method,\cite{book_Golub_Van_Loan:1996} none of the three
methods was better than the others in reducing the nonunitarity. Since both in
the NaI and pyrazine models, only $2\times2$ matrices are relevant, and
since for such matrices, the Jacobi method yields already after one
iteration the analytically exact result for the exponential, we used the
Jacobi method for all results in Sec.~\ref{sec:results}. Note, however, that
the other two methods (based on Pad\'{e} approximants or QR decomposition),
while not exact in the two models used in this paper, converge, in general, faster than
the Jacobi method, and are, therefore, preferred in systems with more than two coupled
electronic states.

\subsection{\label{subsec:trap_midp_algorithms}Trapezoidal rule and implicit
midpoint method}

In addition to nonconservation of energy, the main disadvantage of the
split-operator algorithms is that they can be applied to nonadiabatic dynamics
only in the diabatic representation. Yet, there exist closely related,
arbitrary-order geometric integrators, discussed in Paper I, which, in
addition, conserve energy and are applicable both in the diabatic and
adiabatic representations. These integrators are, like the higher-order
split-operator algorithms, based on recursive symmetric composition (see
Sec.~\ref{subsec:composition}) of the second-order trapezoidal rule
(Crank-Nicolson method\cite{Crank_Nicolson:1947,McCullough_Wyatt:1971}) or the
implicit midpoint method, both of which are, themselves, compositions of the
explicit and implicit Euler methods [see Eqs.~(18), (19), (13), and (14) of
Paper I]. Due to the presence of implicit steps, the trapezoidal rule,
implicit midpoint method as well as their compositions require solving large,
although sparse, linear systems iteratively,\cite{Choi_Vanicek:2019} and, as a
result, in the diabatic representation are expected to be significantly less
efficient than the explicit split-operator algorithms of the same order of
accuracy. These integrators are, again, most naturally implemented in
conjunction with the dynamic Fourier method described in
Sec.~\ref{sec:dyn_fourier}; the only difference being that one must evaluate
the operation $(\hat{T}+\hat{V})\psi$ instead of $\hat{U}_{\hat{T}}\psi$ and
$\hat{U}_{\hat{V}}\psi$. More details about these higher-order integrators can
be found in Paper I,\cite{Choi_Vanicek:2019} which discusses their geometric
properties and studies their efficiency in applications to nonadiabatic
quantum dynamics in the adiabatic representation, in which the molecular
Hamiltonian is nonseparable.

\section{\label{sec:results}Numerical examples}

To test the geometric and convergence properties of the split-operator
algorithms presented in Sections~\ref{subsec:lossprop}%
--\ref{subsec:composition}, we used these integrators to simulate the
nonadiabatic quantum dynamics in a one- and three-dimensional systems.

\subsection{\label{subsec:NaI}One-dimensional model of NaI}

This model is a diabatized version of the one presented in Paper I, i.e., a
one-dimensional two-surface model\cite{Engel_Metiu:1989} of the NaI molecule.
We used the same initial and final times, and the same approximations for the
initial state and for the molecule-field interactions as in Paper I. For
detailed calculation parameters, see Section~III of Ref.~\onlinecite{Choi_Vanicek:2019}.

\begin{figure}
[tbh]\includegraphics[scale=1.0]{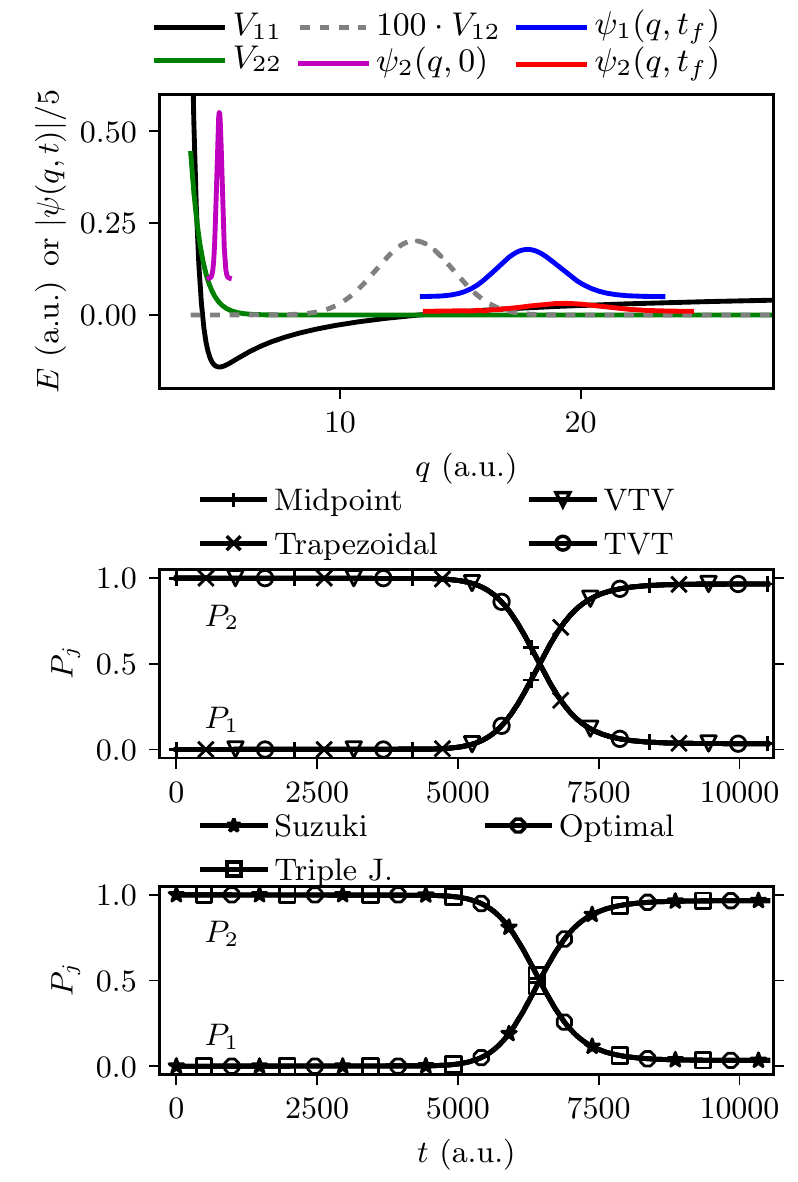}
\caption{\label{fig:P_and_psi}Nonadiabatic dynamics of NaI. Top: Diabatic
potential
energy surfaces with the initial and final nuclear wavepacket components in the
two
diabatic electronic states (the inital ground-state component
is not shown because
it was zero: $\psi_{1}(q,0)=0$).
Middle: Populations of NaI in the two diabatic states computed with four
different second-order methods.
Bottom: Populations computed with three different sixth-order compositions of
the VTV
algorithm.
Populations were propagated
with a time step $\Delta t = 0.01$ a.u. for the second-order methods and $\Delta
t = 82.03125$~a.u. for the
sixth-order methods, i.e., much more frequently than the markers suggest. The
time step guaranteed wavepacket convergence errors below $\approx 10^{-5}$ in
all methods.}
\end{figure}

The top panel of Fig.~\ref{fig:P_and_psi} shows the two diabatic potential
energy surfaces as well as the initial wavepacket at $t=0$ and the ground- and
excited-state components of the final wavepacket at the final time
$t_{f}=10500$ a.u. The population dynamics of NaI, displayed in the middle and
bottom panels of Fig.~\ref{fig:P_and_psi}, shows that after passing this
crossing, most of the population jumps to the other diabatic state, while a
small fraction remains in the original, dissociative diabatic state. On the
scale visible in the figure, the converged populations obtained with the VTV
and TVT split-operator algorithms agree with each other and also with the
results of the trapezoidal and midpoint rule (middle panel). Moreover, the
results of the triple-jump, Suzuki-fractal, and optimal compositions of the
second-order VTV algorithm agree with each other (bottom panel).

\begin{figure}
[pbh]%
\includegraphics[scale=1]{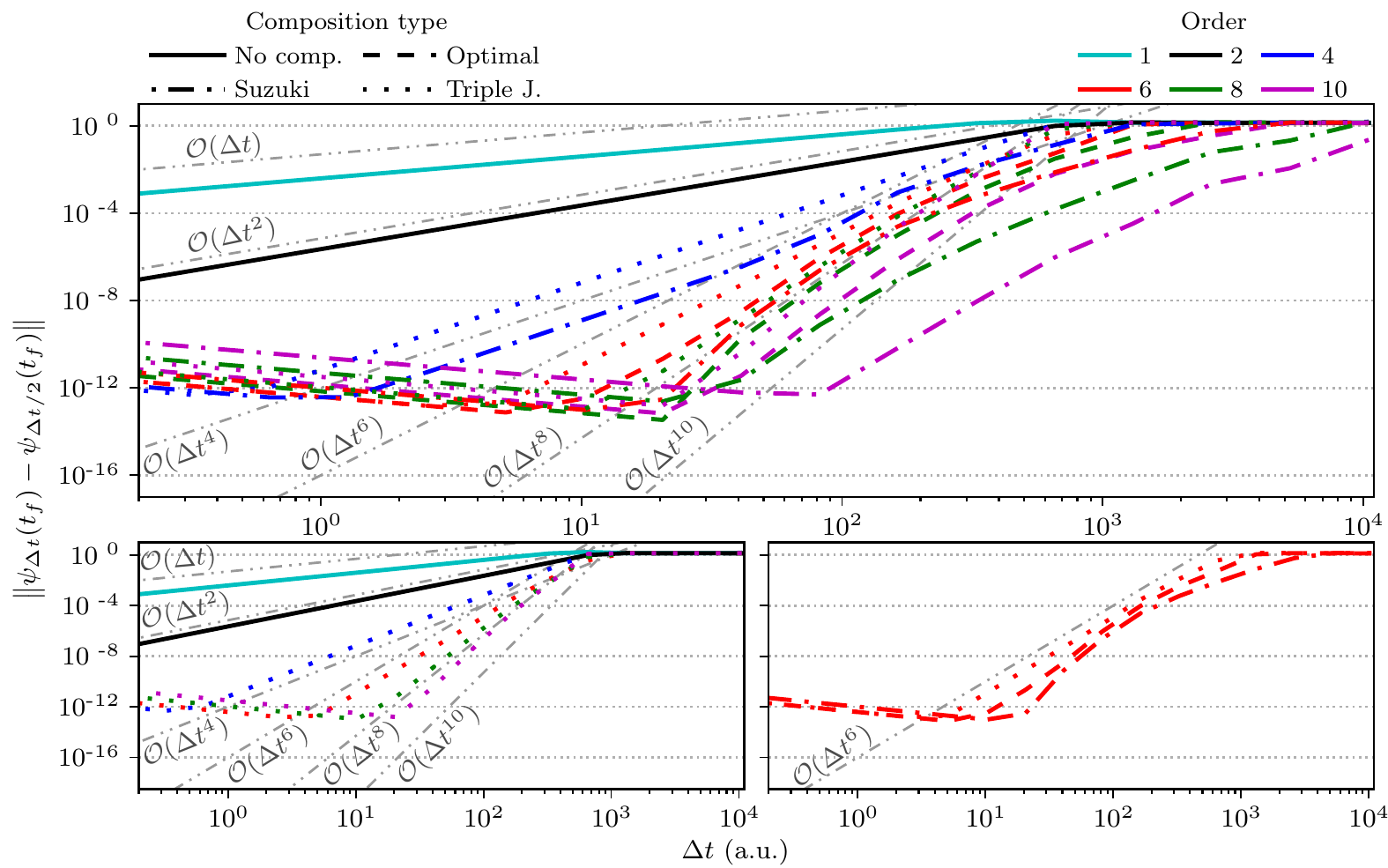}\caption{Convergence of the
molecular wavefunction as a function of the time step. The wavefunction was
propagated with the VT algorithm or with the compositions of the VTV
algorithm. Gray straight lines indicate various predicted orders of
convergence ${\cal{O}}(\Delta t ^{n})$. Top: all discussed methods, bottom left: 
methods composed with
the triple-jump scheme, bottom right: sixth-order methods. }\label{fig:convergence}%

\end{figure}

For a quantitative comparison of various algorithms, it is necessary to
compare their convergence errors at the final time $t_{f}$. As in Paper I, the
convergence error at time $t_{f}$ as a function of the time step $\Delta t$ is
measured by the $L_{2}$-norm error $\left\Vert \psi_{\Delta t}(t_{f}%
)-\psi_{\Delta t/2}(t_{f})\right\Vert $, where $\psi_{\tau}(t_{f})$ represents
the wavepacket propagated with a time step $\tau$. This error is shown in
Fig.~\ref{fig:convergence}, which confirms, for each algorithm, the asymptotic
order of convergence predicted in Secs.~\ref{subsec:lossprop}%
--\ref{subsec:composition}. For clarity, in this and all remaining figures,
only the VT algorithm and compositions of the VTV algorithm are compared
because the corresponding results of the TV algorithm and compositions of the
TVT\ algorithms behave similarly. The top panel of Fig.~\ref{fig:convergence}
compares all methods, whereas the bottom left-hand panel compares only the
different orders of the triple-jump composition and the bottom right-hand
panel compares only different composition schemes with the sixth-order
convergence. Similarly to the results in the adiabatic
basis,\cite{Choi_Vanicek:2019} the prefactor of the error is the largest for
the triple-jump, \cite{Yoshida:1990, Suzuki:1990} intermediate for the
optimal,\cite{Kahan_Li:1997} and smallest for Suzuki-fractal composition. The
figure also shows that for the smallest time steps, the error starts to
increase again. This is due to the accumulating numerical error of the fast
Fourier transform, which eventually outweighs the error due to time
discretization. As a result, the predicted asymptotic order of convergence
cannot be observed for some methods because it is only reached for very small
time steps.

\begin{figure}
[pbh]\includegraphics[scale=1]{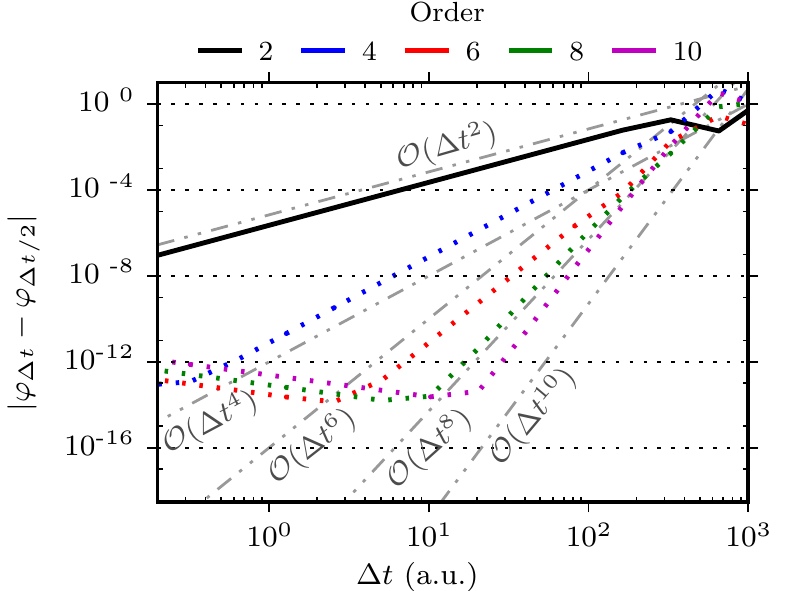}
\caption{Convergence error of the phase of the wavepacket as a
function
of the time step for the triple-jump compositions.
Gray straight lines indicate various predicted orders of
convergence ${\cal{O}}(\Delta t ^{n})$.}\label{fig:phase}
\end{figure}

While the probability density has a classical analogue, the phase of the
wavefunction is a purely quantum property. As a consequence, an accurate
evaluation of the phase is very important in the calculation of electronic
spectra and in other situations, where quantum effects play a role. To
investigate the convergence of the phase as a function of the time step, we
used the phase of wavefunction at the maximum of the probability density (for
a precise definition, see Paper I). Figure~\ref{fig:phase} displays the
convergence of the error of the phase for the triple-jump compositions, and
confirms that the order of convergence is the same as for the wavefunction
itself (bottom left-hand panel of Fig.~\ref{fig:convergence}).

\begin{figure}
[pbh]%
\includegraphics[scale=1]{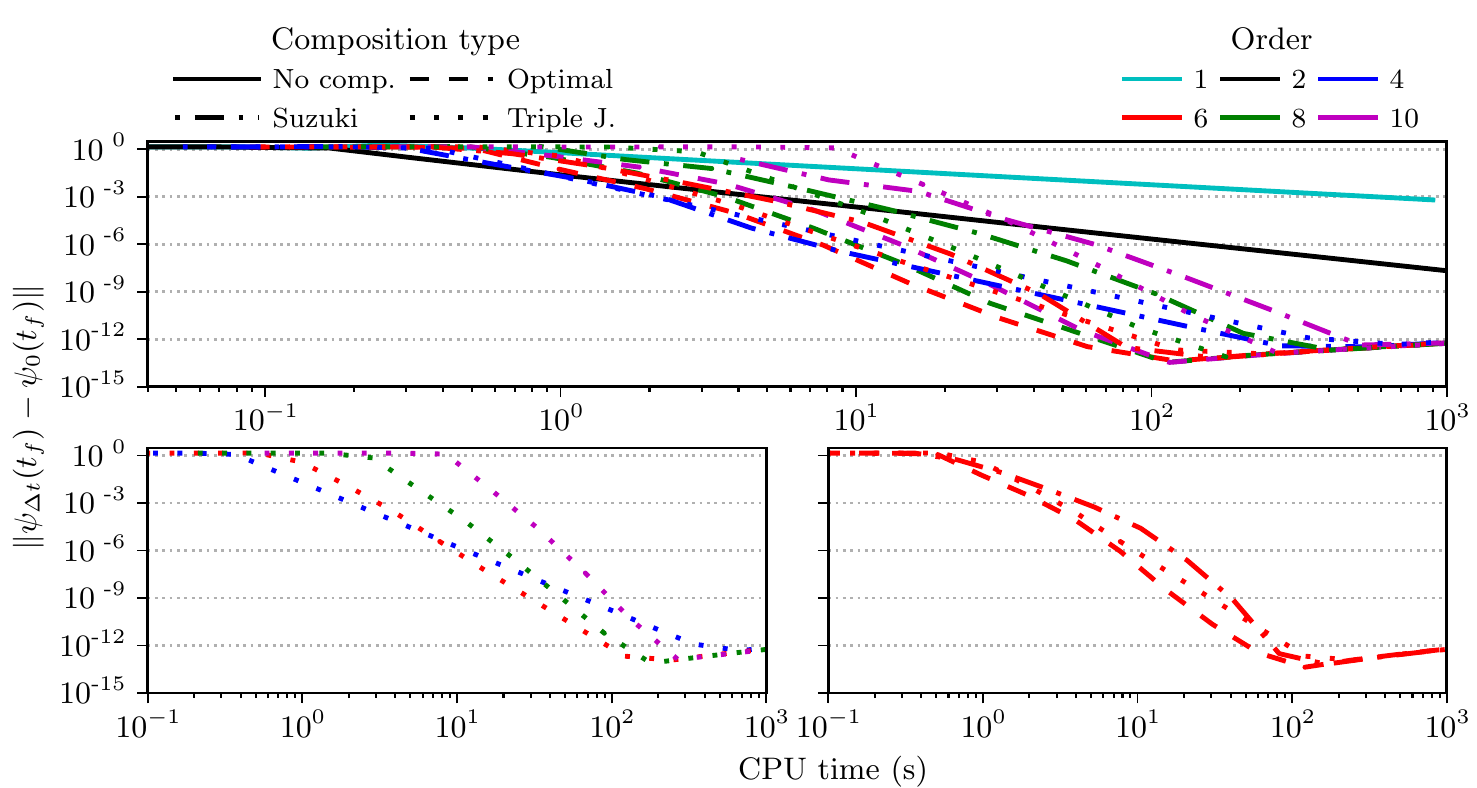}\caption{Efficiency of the
VT algorithm and of various compositions of the VTV algorithm shown using the
dependence of the convergence error on the computational (CPU) time. Top: all
methods, bottom left: triple-jump compositions, bottom right: sixth-order
methods. The reference wavefunction $\psi_{0}(t_{f})$ was chosen as the most
accurate point in Fig. \ref{fig:convergence}, i.e., the wavefunction obtained
using the optimal eighth-order composition with a time step $\Delta
t=t_{f}/2^{9}$.}\label{fig:error_vs_CPU_time}
\end{figure}

Because the number of composition steps depends on the composition scheme and
increases with the order, the efficiency of an algorithm is not determined
solely by the convergence error for a given time step $\Delta t$. It is,
therefore, essential to compare directly the efficiency of the different
algorithms. Figure~\ref{fig:error_vs_CPU_time} displays the wavefunction
convergence error of each algorithm as a function of the computational (CPU)
time. Comparison of the compositions of the VTV split-operator algorithm in
the top panel of Fig.~\ref{fig:error_vs_CPU_time} shows that the fourth-order
Suzuki composition already takes less CPU time to achieve convergence error
$10^{-2}$ than does the elementary VTV algorithm. To reach errors below
$10^{-2}$, it is more efficient to use some of the fourth or higher-order
integrators. Remarkably, the CPU time required to reach an error of $10^{-10}$
is roughly $600$ times longer for the basic VTV\ algorithm than for its
optimal $6^{\mathrm{th}}$-order composition. The bottom right-hand panel of
Fig.~\ref{fig:error_vs_CPU_time} confirms the prediction that the optimal
compositions are the most efficient among composition methods of the same order.

\begin{figure}
[pbh]\includegraphics[scale=1]{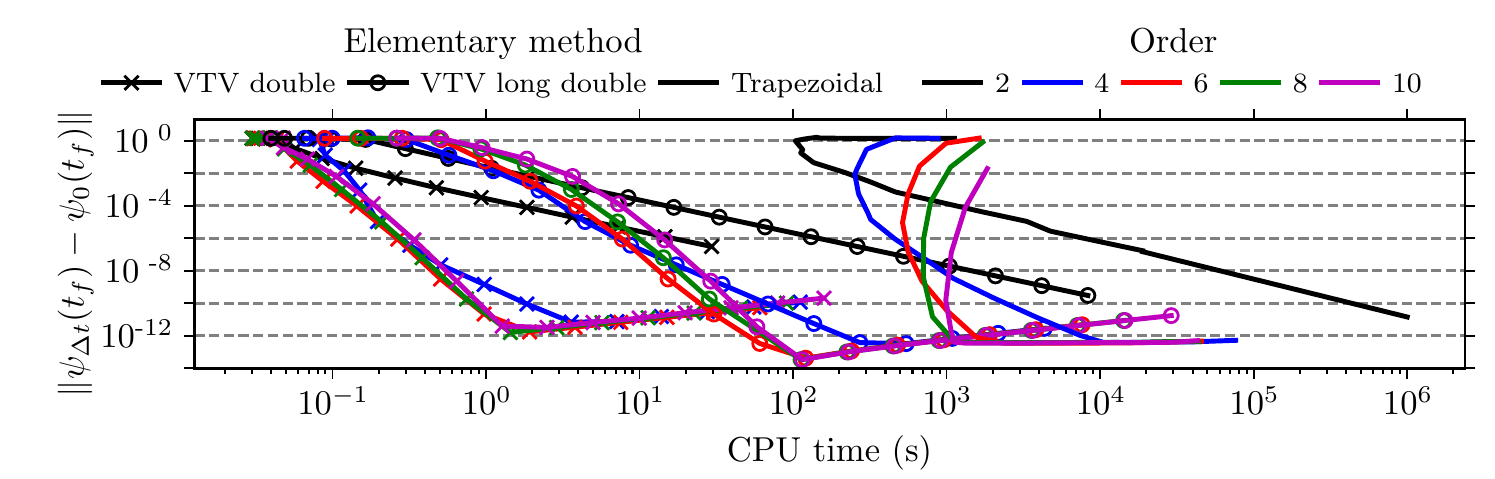}\caption{Efficiency of the
optimal compositions of the trapezoidal rule and of the VTV split-operator
algorithm applied to the NaI model. For the trapezoidal rule, only the double
precision version of the FFTW3 fast Fourier transform was used, while for the
VTV split-operator algorithm, both double and long-double precision versions
are compared. The ``exact'' reference wavefunction $\psi_{0}(t_{f})$ is the
same as in Fig.~\ref{fig:error_vs_CPU_time}. The result of the elementary
second-order trapezoidal rule was extrapolated below the error of
$\approx10^{-7}$ using the line of best fit. As for the fourth-order algorithms,
Suzuki's fractal is considered as the \textquotedblleft optimal
\textquotedblright composition scheme. }\label{fig:efficiency}
\end{figure}

Convergence curves in Figs.~\ref{fig:convergence}--\ref{fig:error_vs_CPU_time}
were obtained using the long-double precision for the FFTW3 algorithm, which
lowered the error accumulation resulting from the nonunitarity of the FFTW3
Fourier transform. If high accuracy is not desired, the double precision of
the FFTW3 algorithm can be used instead, resulting in much more efficient
higher-order algorithms. This is shown for the NaI model in
Fig.~\ref{fig:efficiency}, which compares the efficiency of the optimal
compositions of the VTV algorithm evaluated either with the double or
long-double implementation of the FFTW3, and also with the corresponding
compositions of the trapezoidal rule (for which the double precision of FFTW3
was sufficient). Even the more expensive, long-double precision calculation
with the compositions of VTV algorithm are faster than the corresponding
double precision calculations with the trapezoidal rule, which requires an
expensive iterative solution of a system of linear equations. In particular,
the sixth-order optimal composition of the VTV\ algorithm reaches a
convergence error of $10^{-10}$ forty times faster than the same composition
of the trapezoidal rule (see Fig.~\ref{fig:efficiency}) and $30000$ times
faster than the elementary trapezoidal rule (see
Figs.~\ref{fig:error_vs_CPU_time} and \ref{fig:efficiency}).

Note that the dependence of CPU\ time on the error in
Fig.~\ref{fig:efficiency} is not monotonous for the compositions of the
trapezoidal rule because the convergence of the numerical solution to the
system of linear equations required more iterations for larger time steps; as
a result, both the error and CPU time increased for time steps larger than a
certain critical value.

\begin{figure}
[ptb]\includegraphics[scale=1]{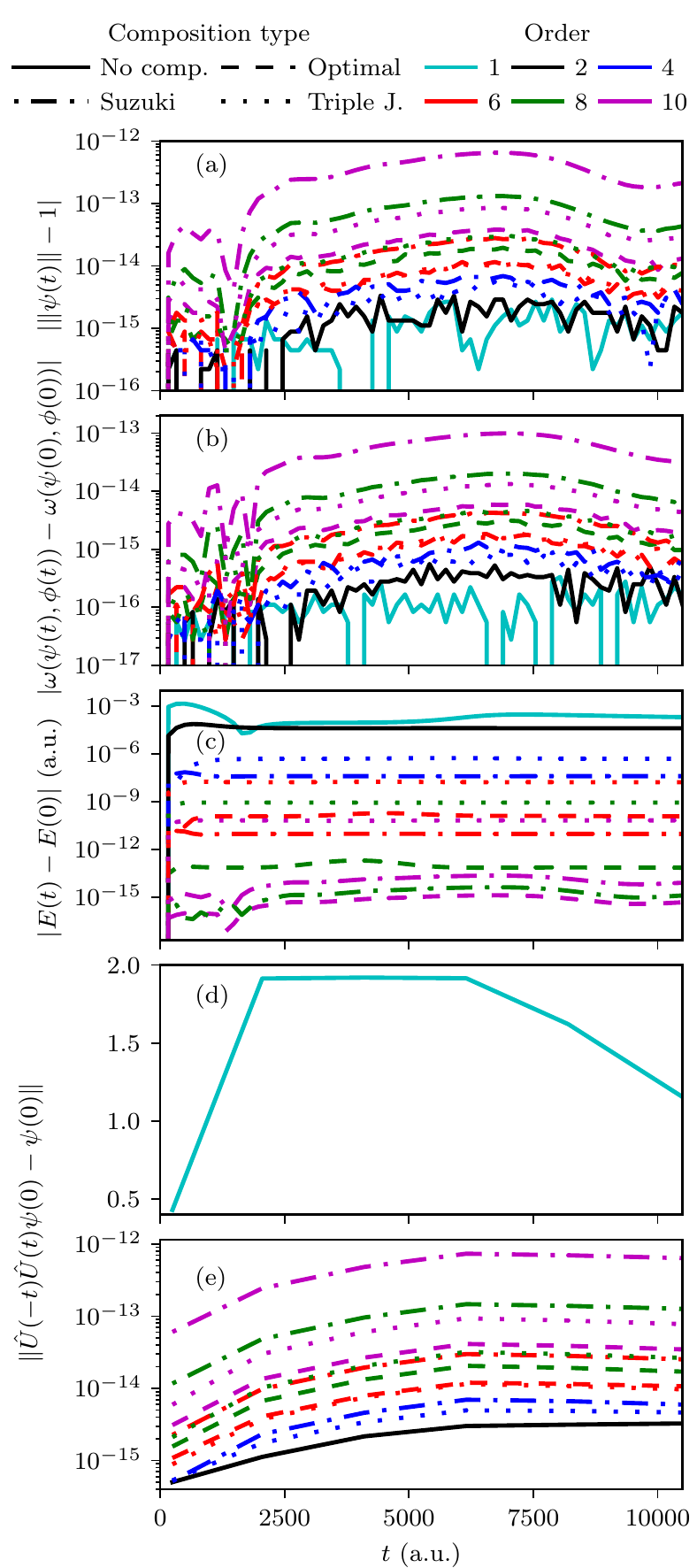}
\caption{Conservation of geometric properties by various algorithms: (a) norm,
(b) symplectic
two-form, (c) energy, and (d)-(e) time reversibility.
$\phi(0)$ is a Gaussian wavepacket with $q_{0} = 5.05$ a.u., $p_{0} = 2.5$ a.u.,
and $\sigma_{0}$ identical
to that of $\psi(0)$. Time reversibility was measured by
the distance of the initial state $\psi(0)$ from
a forward-backward propagated state, i.e., the state $\psi(0)$ propagated
first forward in time for time $t$ and then
backward in time for time $t$. The NaI model and a time step
$\Delta t
= t_{f}/2^{7}$ a.u. was used
in all calculations.
\label{fig:geom_prop}}
\end{figure}

To check that the increased efficiency of higher-order compositions is not
achieved by sacrificing the conservation of geometric invariants, we analyzed,
using the NaI model, the conservation of norm, symplectic two-form, energy,
and time reversibility. Conservation of the norm and symplectic two-form, and
nonconservation of energy by all split-operator algorithms is demonstrated in
panels (a)-(c) of Fig.~\ref{fig:geom_prop}. The tiny residual errors
($<10^{-12}$ in all cases) result from accumulated numerical errors of the FFT
and matrix exponentiation (see Sec.~\ref{sec:dyn_fourier}). Panels (d) and (e)
confirm, on one hand, that the first-order split-operator algorithm is not
time-reversible, and, on the other hand, that the second-order VTV algorithm
together with all its compositions are exactly time-reversible; the tiny
residual errors are again due to accumulated numerical errors of the FFT and
matrix exponentiation.

\begin{figure}
[ptb]\includegraphics[scale=1]{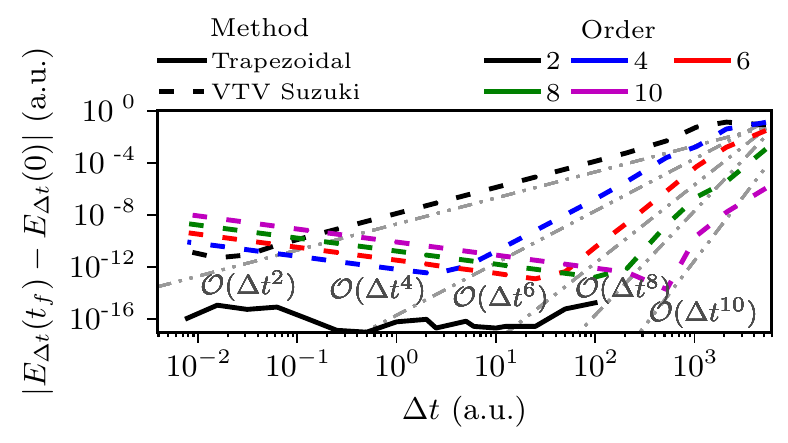}
\caption{\label{fig:E_vs_dt}Energy conservation as a function of the time step
in simulations of the nonadiabatic dynamics of NaI. Gray straight lines indicate 
various orders of
convergence ${\cal{O}}(\Delta t ^{n})$.}
\end{figure}

The nonconservation of energy by the split-operator algorithms is further
inspected in Fig.~\ref{fig:E_vs_dt}, showing the error of energy as a function
of the time step. For the Suzuki-fractal compositions of the VTV algorithm,
the energy is only conserved approximately; its conservation follows the order
of convergence of the integrator, as indicated by the gray lines. In contrast,
the trapezoidal rule conserves the energy to machine accuracy, regardless of
the size of the time step.

\subsection{Three-dimensional model of pyrazine}

\begin{figure}
[pbh]\includegraphics[scale=1]{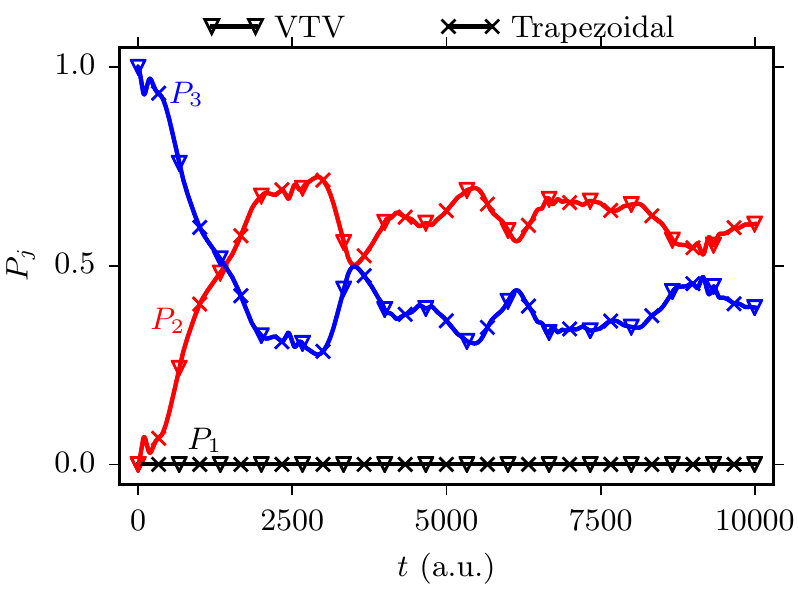}
\caption{Population dynamics of pyrazine obtained using the
sixth-order
optimal compositions of
the trapezoidal rule and VTV algorithm. The same
time
step $\Delta
t=t_{f}/25600$ was used for both
calculations.}\label{fig:pyrazine_pop}
\end{figure}

To investigate how the dimensionality of the system affects the efficiency of
various algorithms, we also performed analogous simulations of a
three-dimensional three-surface vibronic coupling model of pyrazine. The
model, which includes only the normal modes $Q_{1}$, $Q_{6a}$, and $Q_{10a}$,
was constructed by following the procedure from
Ref.~\onlinecite{Stock_Woywod:1995} with the experimental values from
Ref.~\onlinecite{Woywod_Werner:1994} for the vertical excitation energies.
Thirty-two equidistant grid points between $q=-7$~a.u. and $q=7$~a.u. were
included for each vibrational mode. Therefore, the total number of grid points
was increased to $32768$. The initial three-dimensional Gaussian wavepacket
was obtained as the vibrational ground state of the ground-state potential
energy surface ($q_{0}=0$, $p_{0}=0$ and $\sigma_{0}=1$~a.u. for each mode).
Using the sudden approximation, employed also for the NaI model (see Sec.~III
of Paper I), this initial wavepacket was then promoted to the second excited
electronic state and the nonadiabatic quantum dynamics performed until a final
time $t_{f}=10000$~a.u.. The population dynamics, shown in
Fig~\ref{fig:pyrazine_pop}, indicates significant nonadiabatic transitions
between the two excited states, while the ground surface remains unpopulated.
Moreover, on the scale visible in the figure, the population dynamics obtained
with sixth-order optimal compositions of the VTV algorithm and of the
trapezoidal rule agree with each other.

\begin{figure}
[pbh]%
\includegraphics[scale=1]{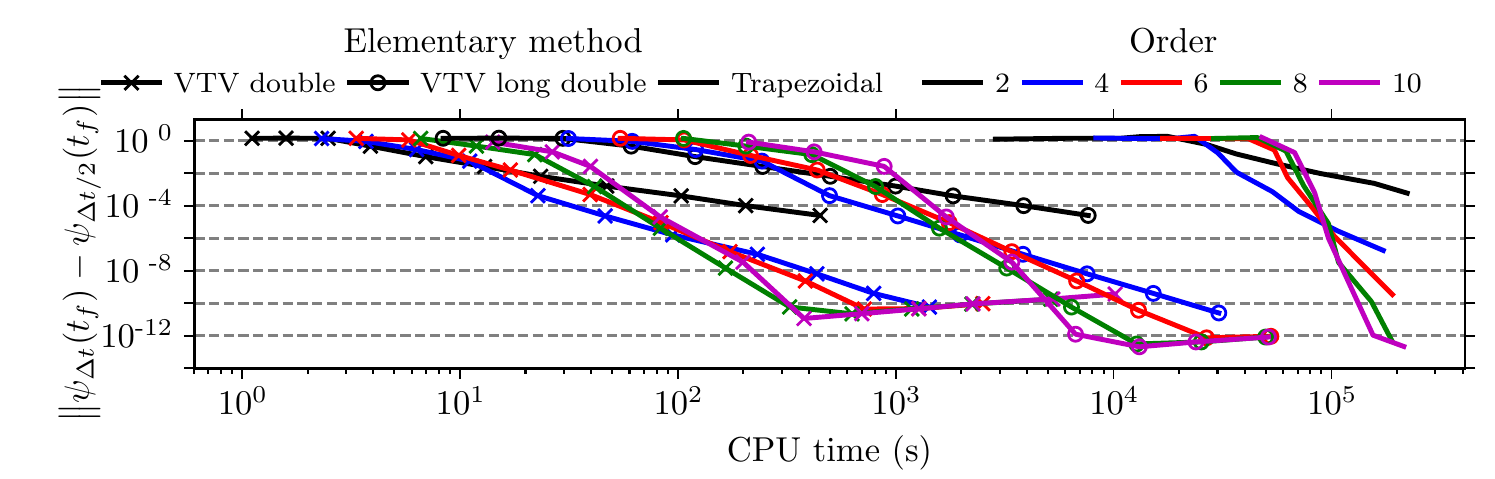}\caption{Efficiency of the
optimal compositions of the trapezoidal rule and VTV split-operator algorithm
applied to the three-dimensional pyrazine model. For the trapezoidal rule,
only the double precision version of the FFTW3 fast Fourier transform was
used, while for the VTV split-operator algorithm, both double and long-double
precision versions are compared. As for the fourth-order algorithms, Suzuki's
fractal is considered as the ``optimal'' composition scheme. }\label{fig:pyrazine_efficiency}%

\end{figure}

Figure~\ref{fig:pyrazine_efficiency} compares the efficiency of different (yet
always optimal) compositions of the VTV algorithm and trapezoidal rule.
Higher-order integrators become more efficient already for convergence errors
below $10^{-2}$ for compositions of the VTV algorithms and, remarkably,
already for errors below $10^{-1}$ for compositions of the trapezoidal rule.
In particular, to reach an error of $10^{-10}$, a $900$-fold speedup over the
second-order VTV algorithm and a $300$-fold speedup over the second-order
trapezoidal rule are achieved by using their tenth-order optimal compositions.
These results suggest that increasing the number of dimensions is either
beneficial or, at the very least, not detrimental to the gain in efficiency
from using the higher-order integrators. As in Fig.~\ref{fig:efficiency}, the
compositions of the VTV algorithms are much more efficient than the
compositions of the trapezoidal rule, but this was expected, because the
Hamiltonian (\ref{eq:diab_mol_ham}) is separable. One must remember that the
main purpose of the compositions of the trapezoidal rule is for nonseparable
Hamiltonians, where the split-operator algorithms cannot be used at all.

\section{\label{sec:conclusion}Conclusion}

We have described geometric integrators for
nonadiabatic quantum dynamics in the diabatic representation, in which the
Hamiltonian is separable into a kinetic term, depending only on
momentum, and potential term, depending only on position. These integrators
are based on recursive symmetric composition of the standard, second-order
split operator algorithm, and as a result, are explicit, unconditionally
stable and exactly unitary, symplectic, symmetric, and time-reversible.
Unlike the
original
split-operator
algorithm, which is only second-order, its recursive symmetric compositions can
achieve accuracy of an arbitrary even order in the time
step. These properties were justified
analytically and demonstrated numerically on a diabatic two-surface model of
NaI
photodissociation. Indeed, the higher-order integrators sped up calculations by
several orders of magnitude when higher accuracy was required. For example, the
computational time required to achieve a convergence error of $10^{-10}$ was
reduced by a factor of $600$ when the optimal sixth-order composition 
was
used instead of the elementary
second-order split-operator algorithm. The gain in efficiency due to the
higher-order integrators was also confirmed by the nonadiabatic simulations in
a diabatic three-dimensional three-surface model of pyrazine. Although
other efficient propagation methods, such as
Chebyshev\cite{Tal-Ezer_Kosloff:1984} or short
iterative Lanczos
schemes,\cite{Lanczos:1950, Park_Light:1986} might have comparable efficiency
in this and other typical chemical systems, in contrast to the integrators
presented here, those methods do not preserve time reversibility and several
other geometric properties of the exact solution.

The authors acknowledge the financial support from the European Research
Council (ERC) under the European Union's Horizon 2020 research and innovation
programme (grant agreement No. 683069 -- MOLEQULE), and from the Swiss
National Science Foundation within the National Center of Competence in
Research \textquotedblleft Molecular Ultrafast Science and
Technology\textquotedblright\ (MUST).

\appendix

\section{\label{appendixa}Geometric properties of numerical integrators}

To simplify many expressions, we set
$\hbar=1$ and denote the increment $\Delta t$ with $\epsilon$ throughout this
appendix. The $\hbar$ can be reintroduced by replacing each occurrence of $t$
with $t/\hbar$ (and $\epsilon$ with $\epsilon/\hbar$). To analyze geometric
properties of various integrators, we will use several 
well-known
identities satisfied by the Hermitian adjoint and inverse operators, listed in
Eqs.~(A1)--(A4) of Paper I.\cite{Choi_Vanicek:2019}

\subsection{\label{localerror}Local error}

The local error of an approximate evolution operator, defined as $\hat
{U}_{\mathrm{appr}}(\epsilon)-\hat{U}(\epsilon)$, is typically analyzed by
comparing the Taylor expansion of $\hat{U}_{\mathrm{appr}}(\epsilon)$ with the
Taylor expansion of the exact evolution operator:%
\begin{equation}
\hat{U}(\epsilon)=1-i\epsilon(\hat{T}+\hat{V})-\frac{1}{2}\epsilon^{2}(\hat
{T}+\hat{V})^{2}+\mathcal{O}(\epsilon^{3})
\label{eq:exact_evol_op_taylor_expansion}%
\end{equation}
If the local error is $\mathcal{O(\epsilon}^{n+1})$, the method is said to be
of order $n$ because the global error for a finite time $t=P\epsilon$ is
$\mathcal{O}(\epsilon^{n})$.

The Taylor expansion of the TV algorithm (\ref{eq:so_TV}) is%
\begin{align}
\hat{U}_{\text{TV}}(\epsilon)  &  =\left(  1-i\epsilon\hat{T}-\frac{1}%
{2!}\epsilon^{2}\hat{T}^{2}\right)  \left(  1-i\epsilon\hat{V}-\frac{1}%
{2!}\epsilon^{2}\hat{V}^{2}\right)  +\mathcal{O}(\epsilon^{3})\nonumber\\
&  =1-i\epsilon(\hat{T}+\hat{V})-\frac{1}{2}\epsilon^{2}(\hat{T}^{2}+2\hat
{T}\hat{V}+\hat{V}^{2})+\mathcal{O}(\epsilon^{3})\nonumber\\
&  =\hat{U}(\epsilon)+\frac{1}{2}\epsilon^{2}[\hat{V},\hat{T}]+\mathcal{O}%
(\epsilon^{3}), \label{eq:taylor_TV}%
\end{align}
so the leading order local error is $\epsilon^{2}[\hat{V},\hat{T}]/2$.
Likewise, for the VT algorithm (\ref{eq:so_VT}),%
\begin{equation}
\hat{U}_{\text{VT}}(\epsilon)=\hat{U}(\epsilon)-\frac{1}{2}\epsilon^{2}%
[\hat{V},\hat{T}]+\mathcal{O}(\epsilon^{3}). \label{eq:taylor_VT}%
\end{equation}

The Taylor expansions of the second-order TVT and VTV\ algorithms are obtained
by composing Taylor expansions~(\ref{eq:taylor_TV}) and (\ref{eq:taylor_VT})
for time steps $\epsilon/2$:
\begin{align}
\hat{U}_{\text{TVT}}(\epsilon)  &  =\hat{U}_{\text{VTV}}(\epsilon)=\hat
{U}\left(  \frac{\epsilon}{2}\right)  \hat{U}\left(  \frac{\epsilon}%
{2}\right)  +\frac{1}{8}\epsilon^{2}\left(  [\hat{V},\hat{T}]-[\hat{V},\hat
{T}]\right)  +\mathcal{O}(\epsilon^{3})\nonumber\\
&  =\hat{U}\left(  \epsilon\right)  +\mathcal{O}(\epsilon^{3}),
\label{eq:trap_rule_imp_mid_taylor_expansion}%
\end{align}
demonstrating that both TVT and VTV are second-order algorithms.

\subsection{\label{subsec:unitarity_symplecticity}Unitarity, symplecticity,
and stability}

Both first-order split-operator algorithms are unitary because%
\begin{align*}
\hat{U}_{\text{TV}}(\epsilon)^{-1}  &  =e^{i\epsilon{\hat{V}}}e^{i\epsilon
{\hat{T}}}=\hat{U}_{\text{TV}}(\epsilon)^{\dag},\\
\hat{U}_{\text{VT}}(\epsilon)^{-1}  &  =e^{i\epsilon{\hat{T}}}e^{i\epsilon
{\hat{V}}}=\hat{U}_{\text{VT}}(\epsilon)^{\dag}.
\end{align*}
Both second-order split-operator algorithms are unitary because they are
compositions of unitary first-order algorithms.

Because the symplectic form was defined in Sec.~\ref{subsec:exactprop} as the
imaginary part of the inner product and because VT, TV, VTV, and TVT
algorithms as well as their compositions are unitary, all of them are also symplectic.

Stability follows from unitarity because
\begin{equation}
\Vert\psi(t+\epsilon)-\phi(t+\epsilon)\Vert=\Vert\psi(t)-\phi(t)\Vert
\label{eq:trap_rule_imp_mid_stable}%
\end{equation}
for unitary evolution operator $\hat{U}_{\text{appr}}(\epsilon)$. Since all
split-operator methods are unitary, all are stable as well.

\subsection{\label{subsec:comm_energy_cons}Commutation of the evolution
operator with the Hamiltonian and conservation of energy}

Because the kinetic and potential energy operators do not commute, unless
$\hat{V}=\operatorname*{const},$ the evolution operator of no split-operator
algorithm commutes with the Hamiltonian. E.g., for the TV algorithm,
\begin{equation}
\lbrack\hat{H},\hat{U}_{\text{TV}}(\epsilon)]=[\hat{T}+\hat{V},e^{-i\epsilon
{\hat{T}}}e^{-i\epsilon{\hat{V}}}]=e^{-i\epsilon{\hat{T}}}[\hat{T}%
,e^{-i\epsilon{\hat{V}}}]+[\hat{V},e^{-i\epsilon{\hat{T}}}]e^{-i\epsilon
{\hat{V}}}\neq0. \label{eq:TV_comm}%
\end{equation}
As a consequence, split-operator algorithms do not conserve energy.

\subsection{\label{subsec:symm_revers}Symmetry and time reversibility}

As shown, e.g., in 
Refs.~\onlinecite{book_Leimkuhler_Reich:2004,book_Hairer_Wanner:2006} or in
Appendix A of Paper I, the adjoint of an evolution operator satisfies the
following properties:%
\begin{align}
(\hat{U}(\epsilon)^{\ast})^{\ast}  &  =\hat{U}(\epsilon),\label{eq:adj_adj}\\
(\hat{U}_{1}(\epsilon)\hat{U}_{2}(\epsilon))^{\ast}  &  =\hat{U}_{2}%
(\epsilon)^{\ast}\hat{U}_{1}(\epsilon)^{\ast},\label{eq:adj_prod}\\
&  \hat{U}(\epsilon)\hat{U}(\epsilon)^{\ast}\text{ is symmetric.}
\label{eq:sym_op_w_adj}%
\end{align}
Note that the third property gives a simple recipe for developing symmetric
methods---by composing an arbitrary method and its adjoint, with both
composition coefficients of $1/2$.

The first-order VT and TV split-operator algorithms are adjoints of each other
because
\begin{equation}
\hat{U}_{\mathrm{TV}}(-\epsilon)^{-1}=e^{-i\epsilon{\hat{V}}}e^{-i\epsilon
{\hat{T}}}=\hat{U}_{\mathrm{VT}}(\epsilon) \label{eq:exp_imp_eu_adjoint}%
\end{equation}
and because of Eq.~(\ref{eq:adj_adj}). Therefore, neither VT\ or TV algorithm
is symmetric or time-reversible. In contrast, the second-order VTV and TVT
algorithms are both symmetric, which follows from Eq.~(\ref{eq:sym_op_w_adj})
applied to the two possible compositions of the VT and TV algorithms with
composition coefficients $1/2$. As shown, e.g., in 
Refs.~\onlinecite{book_Leimkuhler_Reich:2004, book_Hairer_Wanner:2006} or in
Appendix A of Paper I, time reversibility follows from symmetry. Therefore,
both VTV and TVT algorithms and their symmetric compositions are time-reversible.

\section{\label{appendixb}Exponential convergence with grid density}

\begin{figure}
[tbh]\includegraphics[scale=1.0]{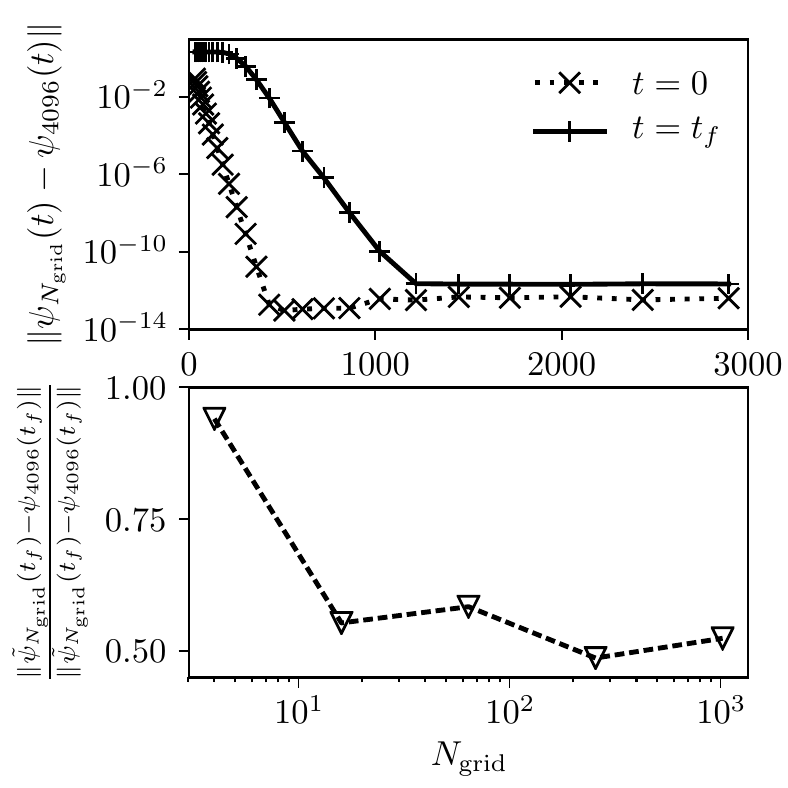}
\caption{Top: Convergence of the initial and final wavepackets
with the increasing
number of grid points. Bottom: Ratio of the integration error and total convergence error at the final time
as a function of the
number of grid points. (See Appendix~\ref{appendixb}  for details.)
The sixth-order optimal composition of the VTV
algorithm with
time step $\Delta t=
t_{f}/2^{7}$ was used for propagation.\label{fig:gridconv}}
\end{figure}

The top panel of Fig.~\ref{fig:gridconv} exhibits the
exponential convergence of the molecular wavefunction with the increasing
number of grid points for the NaI model in the diabatic basis. The ranges as
well as the densities of both the position and momentum grids were increased
by a factor of $\sqrt{2}$ for each increase in the number $N_{\text{grid}}$ of
grid points by a factor of two. Convergence error required comparing
wavefunctions on grids with different densities, which was carried out by
trigonometric interpolation of the wavefunction on the sparser grid.
Increasing $N_{\text{grid}}$ reduces the 
convergence error at
time $t_{f}$ (top panel) because the errors of both the 
required
overlap integral and of the propagation decrease. To compare these two
effects, the bottom panel of Fig.~\ref{fig:gridconv} shows the ratio of the
purely integration error and the total error. The integration error is defined
as $\Vert\tilde{\psi}_{N_{\mathrm{grid}}}(t_{f})-\psi_{4096}(t_{f})\Vert$
where $\psi_{4096}(t_{f})$ is the wavefunction propagated on the fully
converged grid and $\tilde{\psi}_{N_{\mathrm{grid}}}(t_{f})$ is $\psi
_{4096}(t_{f})$ represented with $N_{\mathrm{grid}}$ grid points. In other
words, the representation on a reduced grid is done only after propagation.
The panel shows that at the final time, the integration error is approximately
one half of the total error. Therefore, the integration and propagation errors
due to a finite grid are similar.

\newpage
\bibliographystyle{aipnum4-1}
\bibliography{integrators_nonadiabatic}

\end{document}